# Exceptionally high, strongly temperature dependent, spin Hall conductivity of SrRuO₃

Yongxi Ou[1,2*†], Zhe Wang[1†], Celesta S. Chang[1,2†], Hari P. Nair[3], Hanjong Paik[3,4], Neal Reynolds[2], D.C. Ralph[2,5], D. A. Muller[1,5], D. G. Schlom[3,5] and R. A. Buhrman[1*]

[1]School of Applied and Engineering Physics, Cornell University, Ithaca, New York 14853, USA,

[2]Department of Physics, Cornell University, Ithaca, New York 14853, USA,

[3]Department of Materials Science and Engineering, Cornell University, Ithaca, New York 14853, USA,

[4]Platform for the Accelerated Realization, Analysis, & Discovery of Interface Materials (PARADIM), Cornell University, Ithaca, New York 14853, USA,

[5]Kavli Institute at Cornell for Nanoscale Science, Ithaca, New York 14853, USA

*email: yo84@cornell.edu; rab8@cornell.edu.

†: These authors contributed equally to this work.

## Abstract

Spin-orbit torques (SOT) in thin film heterostructures originate from strong spin-orbit interactions (SOI) that, in the bulk, generate a spin current as the result of extrinsic spin-dependent, skew or/and side-jump, scattering, or in the intrinsic case due to Berry curvature in the conduction band. While most SOT studies have focused on materials with heavy metal components, the oxide perovskite SrRuO₃ has been predicted to have a pronounced Berry curvature. Through quantification of its spin current by the SOT exerted on an adjacent Co ferromagnetic layer, we determine that SrRuO₃ has a strongly temperature (*T*) dependent spin Hall conductivity which becomes particularly high at low *T*, e.g. $\sigma_{SH} \geq (\hbar/2e) 3 \times 10^5 \ \Omega^{-1} \mathrm{m}^{-1}$ at 60 K. Below the SrRuO₃ ferromagnetic transition, non-standard SOT components develop associated with the magnetic characteristics of the oxide, but these do not dominate as with spin currents from a conventional ferromagnet. Our results establish a new approach for the study of SOI in epitaxial conducting oxide heterostructures and confirm SrRuO₃ as a promising candidate material for achieving new and enhanced spintronics functionalities.

Strong spin-orbit interactions (SOI) in conducting systems have long been a subject of keen fundamental interest and practical importance, accounting, e.g. for the source of the anomalous Hall effect (AHE) in magnetic conductors [1–3]. Until recently the direct experimental study of these effects on electron and spin conductivity has been largely limited to magneto-optical studies [4,5] and to measurements of the AHE, where the effect of the SOI on electron transport is intertwined with the magnetic properties [1]. The recent development of methods for quantifying the transverse *spin Hall conductivity* $\sigma_{SH}$ of conducting systems through measurements of the "spin-orbit torques" (SOT) [6–10] that the spin current exerts when it impinges on an adjacent ferromagnetic layer, has established a powerful new technique for studying strong spin-orbit interactions in non-magnetic conductors with high $\sigma_{SH}$. So far, the search for materials that can provide robust SOT has concentrated on conductors where at least one component is a heavy metal (HM) element, e.g. Pt, W, Bi, Ir [11–18], but it is known strong SOI can also be found in materials composed of only lighter elements. In particular the Ru ion is understood to be the origin of a quite strong spin-orbit interaction in the conduction bands of the ruthenate class of oxide conductors [19–21], $Sr_{n+1}Ru_nO_{3n+1}$, accounting for the very strong magneto-crystalline anisotropy (MCA) of $SrRuO_3$ ($n=\infty$) in its low temperature ferromagnetic state [19]. Band structure calculations have indicated a particularly strong Berry curvature in this material [22,23], which is now understood to be a good predictor of a high intrinsic $\sigma_{SH}$ (strong spin Hall effect), although AHE measurements have not fully supported a dominant role of the Berry curvature in that aspect of ferromagnetic $SrRuO_3$ [24].

Here we report SOT and related spin transport measurements of high quality $SrRuO_3$/Co thin film bilayers grown *in situ* in a molecular beam epitaxy (MBE) system where the highly ordered $SrRuO_3$ (hereafter SRO) is grown on a $SrTiO_3$ (STO) buffer layer formed on a Si substrate. From spin-torque ferromagnetic resonance (ST-FMR) we find that the damping-like SOT that is exerted on the Co layer from RF current flowing in the SRO layer is quite strong at room temperature, and somewhat surprisingly, *increases* quasi-linearly with decreasing temperature to the lowest temperature studied, ~ 60 K, well below the ferromagnetic Curie temperature $T_c$ of the SRO layer of ~ 150 K. Effects of the ferromagnetic transition and the AHE for $T \leq T_c$ are also clearly evidenced in the temperature dependent SOT.

The fabrication of the oxide/FM heterostructures used in our spin-torque study began with the epitaxial growth of first a 15 nm thick STO buffer layer and then the SRO (4, 6 or 10 nm) on a Si (001) substrate, using a recently refined MBE technique [25] (see Supplementary Material), following by the deposition of a thin film of Co (4 or 6 nm thick) on top of SRO with a 1 nm thick Al capping layer. Figure 1a. provides a schematic representation of the epitaxial heterostructure. We confirmed the quality of the SRO layers by X-ray diffraction (see SM) and by scanning transmission electron microscopy (STEM). Figure 1b shows the low magnification angular bright field (ABF) image of the STO(15)/SRO(6)/Co(6) heterostructure (thickness of layers in nm). We also measured the resistivity of a STO(15)/SRO(10) sample as a function of temperature, with results as shown in Fig.1d. The resistivity of the SRO $r_{xx}^{SRO}(T)$ shows a strong temperature-dependent behavior from room temperature down to 4 K, typical of the material [26,27]. There is an obvious kink at ~ 150 K in the resistivity, consistent with a change in electron scattering rate due to the ferromagnetic transition of SRO [26]. In the inset of Fig.1d, we plot $d\, r_{xx}^{SRO}(T)/dT$ as a function of $T$, from which the Curie temperature can be estimated to be $T_c \sim 150K$ , slightly lower than the $T_c$ in bulk value ($\sim 160K$ ), but close to values previously observed in good quality thin SRO films [28]. Figure 1c displays shows high magnification high-angle annular dark-field (HAADF) and ABF STEM images of a SRO(10)/Co(6) sample showing that the Co layer is polycrystalline. As seen clearly in the ABF image, the well-defined atomic positions and clear separation of the layer structures at the SRO/Co interface indicate there is little distortion or intermixing there, which is promising for efficient interfacial spin transport.

In the spin Hall effect (SHE) a transverse spin current density $j_s$ is generated in the direction ($\hat{z}$) normal to the plane of the spin Hall layer with the spin polarization $\hat{s}$ being in-plane ($\hat{s}_y$), orthogonal to the direction ($\hat{x}$) of the electrical current density $j_c$. $\theta_{SH} \equiv (2e/\hbar)j_s/j_c$ quantifies the strength of the SHE while the damping-like spin torque efficiency $\xi_D \equiv T_{int}\theta_{SH}$ as determined for SOT measurements accounts for the less than perfect spin transparency $T_{int}$ of the interface with the adjacent FM SOT detector [29,30]. In ST-FMR (see SM), as long as the field-like component of SOT is much weaker than the torque due to the Oersted field, the damping-like spin torque efficiency $\xi_D$ of the SRO/Co heterostructure can be determined from the ratio of the symmetric (S, $V_S$) and antisymmetric (A, $V_A$) components of the FMR response to an applied RF bias current [8,30]:

$$\xi_D = \frac{V_S}{V_A}\left(\frac{e}{\hbar}\right)\mu_0 M_s t_{Co} d_{SrRuO_3}\sqrt{1+M_{eff}/H_{res}} \qquad (1).$$

Here $e$ is the electron charge, $\hbar$ the reduced Planck constant, $\mu_0$ the vacuum permeability, $M_s$ the saturation magnetization of cobalt, $t_{Co}$ ($d_{SrRuO_3}$) the Co (SRO) thickness, $M_{eff}$ the Co demagnetization field, and $H_{res}$ the resonance field.

Table I summarizes $\xi_D$ as determined by room-temperature (300 K) ST-FMR measurements for a series of SRO($d_{SRO}$)/Co($t_{Co}$) samples. A substantial $\xi_D$ is found for all three thicknesses of SRO, with $\xi_D$ rising to 0.1 for the SRO(10) samples. Note that for a given SRO thickness, $\xi_D$ is almost identical for different Co thicknesses, which indicates that the field-like torque in our SRO/Co is dominated by the Oersted field $H_{Oe}$ arising from the RF current flowing in the SRO, since $V_S / V_A$ would show a nontrivial dependence on $t_{Co}$ if there was a significant field-like torque contribution arising from either an interfacial Rashba-Edelstein effect or from interfacial scattering of the incident spin current from the SRO [30].

Typically, when a thickness-dependent $\xi_D$ is observed in measurements of HM/FM heterostructures, it is attributed to the fact that the maximum spin current density from the SHE is only approached once the HM thickness is $> \lambda_s$, the spin diffusion length [31,32]. This would indicate $l_s^{SRO} > 6$ nm in our samples, which seems unlikely for SRO at room temperature with its "bad metal" short mean free path and apparent strong SOI. Moreover a long diffusion length explanation is not consistent with $\xi_D$ being nearly the same for the SRO(4) and SRO(6) samples.

It is more likely that the SRO electronic structure has a significant dependence on thickness due to a strain effect from the STO/SRO interface that modifies the transport properties of the material, including $\sigma_{SH}$. To examine this possibility, we performed high resolution STEM to quantify the position-dependent tilt of the $RuO_6$ octahedra in SRO films of two different nominal thicknesses, 6 nm and 10 nm. Results are shown in Fig. 2. The octahedral tilt in the bulk of the thicker SRO is above $20°$ while it is only about $8°$ in the thinner SRO sample. A recent theoretical calculation of the spin Hall conductivity of orthorhombic $SrIrO_3$ has concluded that tilting and rotation of the oxygen octahedron is crucial to the large spin Hall effect predicted for that semi-metal pervoskite [33]. We also note that the measured tilt of both SRO samples drops down to ~ 5$^o$ in the last layer adjacent to the Co, which may be indicating a spatially varying spin Hall conductivity near the SRO/CO interface and hence that the spin current density at the SRO/Co interface is less than that in bulk of the SRO.

To examine the effect of the strongly temperature (*T*) dependent electron conductivity of SRO and also the effect of its ferromagnetic transition on the spin current, we performed *T*-dependent ST-FMR measurements on several SRO(10)/Co(6) samples. Due to the possibility, particularly when the SRO is below its Curie temperature, that the spin current from the SRO might have polarization components ($\hat{s}_z$ and/or $\hat{s}_x$) in addition to the standard SHE $\hat{s}_y$ polarization, at each *T* we measured the ST-FMR response as the function of the angle $\varphi$ of the Co magnetization relative to the RF bias current direction $\hat{x}$ by varying the orientation of the in-plane magnetic field bias. For a transverse spin current of arbitrary spin polarization, the dependence of the SOTs on the magnetization direction results in the following for both the S and A components (see SM):

$$V_{S(A)} = V_{S(A)}^y \cos\varphi \sin 2\varphi + V_{S(A)}^z \sin 2\varphi + V_{S(A)}^x \sin\varphi \sin 2\varphi \qquad (2).$$

In Eq.(2), the first term $V_{S(A)}^y$ is determined by the strength of the torques exerted on the FM by the standard SHE-generated spin current ($H_{\text{Oe}}$), which represent the in-plane (//) (out-of- plane ($\perp$)) torque with symmetry $\tau_{//(\perp)}(\varphi) = \tau_{S(A)} \cos\varphi$. The second and third terms denote the ST-FMR responses that arise from the torques exerted by any incident spin currents with, respectively, $\hat{s}_z$ and $\hat{s}_x$ polarization (or from the torques exerted by any effective field that is along $\hat{z}$ or $\hat{x}$). As we discuss below, the signals from our SRO/Co samples are, as expected, dominated by the first term in Eq. (2), but at low $T$ there is also a measurable component whose $\varphi$ dependence is described by the second term. Figure 3a shows the schematics for $\tau_{//}$ and $\tau_\perp$ in the SRO/Co bilayer.

We show in Fig.3b $\xi_D$ as determined from the $V_S^y / V_A^y$ ratio of the ST-FMR response for one of our SRO(10)/Co(6) microstrips as a function of $T$ from 300 K to 60 K. Within this $T$ range, $\xi_D$ increases monotonically and substantially, as $T$ is decreased (a very similar result was obtained with a second sample; see SM). This $\xi_D$ behavior is unusual as the SRO conductivity $S_{xx}(T)$ is also increasing with decreasing $T$ (see the inset of Fig.3c). Since we have $\xi_D = T_{\text{int}} (2e/\hbar) \sigma_{SH} / \sigma_{xx}$, our results indicate a strongly varying $S_{SH}(T)$, increasing by as much as 8 times, from room temperature to 60 K, unless $T_{\text{int}}$ is also strongly $T$-dependent, which we consider unlikely (see SM). With the assumption that $T_{\text{int}}$ is constant, in Fig. 3c we plot the *effective* spin Hall conductivity, $\sigma_{SH}^{\text{eff}}(T) \equiv T_{\text{int}} \sigma_{SH}(T) = (\hbar/2e) \xi_{DL}(T) \sigma_{xx}(T)$ of this SRO(10)/Co(6) bilayer showing that it attains the value of $3.2 \times 10^5 (\hbar/2e) \Omega^{-1} \text{m}^{-1}$ at 60 K, which while exceptionally high for a material with no heavy elements and a low carrier density is only a lower bound since $T_{\text{int}} < 1$.

In addition to the high $S_{SH}^{eff}(T)$ and its strong variation with $T$, another surprising result is the minimal effect of the paramagnetic-to-ferromagnetic transition of the SRO on $\xi_D(j_s)$, (Fig. 3b). Normally one would expect any spin current generated by the SHE in a FM to be rapidly dephased due to precession about the local exchange field, and that only a spin current component whose polarization is collinear with the internal magnetization would be retained [34]. Our tentative interpretation for the persisting strong transverse spin current with $\hat{s}_y$ polarization persists in SRO well below $T_c$ is that this is due to the spin diffusion length $l_s$ in SRO being considerably *shorter* than the spin dephasing length $l_{dp}$ in the FM state. In a conventional ferromagnet, *e.g.* Co, $\lambda_{dp} \simeq 1$ nm $<< l_s$ [32,35], but the lower SRO exchange energy (field) in combination with its very strong spin-orbit interaction (short $\lambda_s$) appears to reverse this relationship resulting in an atypical preservation of a strong spin Hall current below the SRO magnetic transition.

There is clear evidence of the FM state in the spin torque behavior; that is found in the smaller response of the ST-FMR signal that has the symmetry expected for a spin current with $\hat{s}_z$ polarization, as described by the second term in Eq. 2. In that case $V_S^z$ arises from an in-plane field-like spin-orbit torque via a perpendicular effective field $H_{\mathrm{eff}}^z$ and $V_A^z$ is the measure of the out-of-plane damping-like torque generated by the Co absorption of $\hat{s}_z$ spins (see SM). In Fig. 3d and its inset we plot as a function of $T$ the spin torque efficiency $\chi_D^z$ and $H_{\mathrm{eff}}^z$, respectively, due to $\hat{s}_z$. Since our SRO films are twinned, some of the material has its b-axis oriented $45°$ out of plane. This is the easy axis of the very strong MCA of SRO in the FM state, with reported anisotropy fields $> 1$ Tesla [19]. In this case we would expect the spin current arising from the AHE to have, due to the partial out-of-plane magnetization, both an in-plane component and an out-of-plane $\hat{s}_z$ component. We attribute $\chi_D^z$ to the absorption by Co of this AHE generated $\hat{s}_z$ spin current, with the sharp rise in its amplitude just below $T_c$, being due to the rapid development of the exchange field that is required to produce this spin orientation from the underlying SHE. The amplitude of $\chi_D^z$ at 60 K, together with the value of $s_{xx}$, indicates a lower bound (due to the undetermined value of $T_{int}$) for the transverse spin conductivity associated with the $\hat{s}_z$ polarized spin current of $S_{xy}^{s_z} \approx (\hbar/2e) 5 \times 10^3 \ \Omega^{-1} m^{-1}$, which is somewhat less than but of similar magnitude as the AHE conductivity previously reported for SRO films [22,36]. The fact that $S_{xy}^{s_z}$ and $S_{AH}$ are both $<< S_{xy}^{s_y}$ suggests that, since the AHE is approximately the difference of the contributions of Berry curvature from the two spin populations while the SHE is the sum, the conduction electron spin polarization of SRO is low, as indicated previously by some measurements but not all [19].

We estimate $H_{Oe}$ as generated by the RF current in the SRO to be $\approx 6 \times 10^{-11} \ \text{Oe}/(Am^{-2})$. As can be seen in the inset of Fig. 3d, $H_{eff}^z$ is only $\sim 15\%$ of $H_{Oe}$ at 60 K. We tentatively ascribe this small $H_{eff}^z$ to be the effective field that is exerted on the SRO magnetization by the limited precession of the spin Hall current, with this field then acting on the Co. We note that there is a clear magnetic interaction between the SRO and the Co as evidenced by the change in the demagnetization field of Co behavior below $T_c$ of SRO (see SM). Since the dimensions ( $20 \ \mu m \times 10 \ \mu m$ ) of the patterned SRO/Co microstrips used in the ST-FMR measurements are only somewhat larger than the size (~5 μm) of the twinned domains reported in similar SRO layers [37], the details of the $\chi_D^z$ and $H_{eff}^z$ signals, and also the behavior of the Co demagnetization field, should vary somewhat from sample to sample, which is what we observe (see SM). We also analyzed the magnitude of the torques described by the third term in Eq.(2) of the angular dependent ST-FMR signal, but found them either non-existent or much weaker than the other two terms (not shown here).

The $T$-dependence that we observe in $\chi_{DL}$, and hence the stronger variation in $\sigma_{SH}^{\text{eff}}(T)$, is not consistent with a dominant extrinsic origin of the SHE in SRO. The skew-scattering contribution to $\sigma_{SH}(T)$ is expected to scale linearly with the electron lifetime $\tau$ (*i.e.* with $\sigma_{xx}(T)$), which would give a $T$-independent contribution to $q_{SH}$ ($\chi_{DL}$), while the side-jump contribution to $\sigma_{SH}(T)$ is expected to be independent of $\sigma_{xx}(T)$ [38], which predicts $q_{SH} \propto 1/\sigma_{xx}(T)$, *i.e.* a *decrease* with decreasing $T$. For an intrinsic contribution, although $\sigma_{SH}$ should ordinarily be independent of changes in $\tau$ at fixed temperature as long as the band structure is constant, a strong temperature dependence is still possible due to strong $T$-dependent changes in self-energy as evidenced by $\sigma_{xx}(T)$ and/or simply the role of $T$ in modifying the occupation of the nearly-degenerate regions of the conduction bands of SRO where there is strong Berry curvature [39,40]. We note that a first-principles calculation of the intrinsic $\sigma_{SH}(T)$ of Pt [40] has predicted a $T$-dependence that is very similar to what we observe here for SRO. Alternatively, part of the $\sigma_{SH}(T)$ behavior could arise from a $T$-dependent change in the octahedral tilt due to the change in the interfacial strain, but since Co has a slightly higher coefficient of thermal contraction than SRO [41,42], any detrimental strain at the SRO/Co interface should increase with decreasing $T$.

In summary, from ST-FMR measurements we obtain a strong, damping-like SOT attributable to the spin Hall effect in the high-quality SRO/Co bilayers that increases with decreasing temperature and continues to do so well below the Curie temperature of the SRO.

This behavior, in conjunction with the strong temperature dependence of $S_{xx}$, indicates that the intrinsic SRO spin Hall conductivity $S_{SH}(T)$ *increases* strongly with decreasing *T,* reaching as a lower bound $\sigma_{SH} \geq (\hbar/2e) \times 3 \times 10^5 \ \Omega^{-1} m^{-1}$ at 60 K. We suggest that this *T* behavior is due to changes in the electron self-energy and/or changes to the occupation of states in those nearly degenerate regions of the SRO conduction bands where there is strong Berry curvature. We also observed spin torques attributable to a smaller spin current being emitted by the SRO in the FM state with out-of-plane polarization which is comparable to what might be expected from the AHE previously reported for SRO films. Our results open a new avenue for the fundamental study of strong SOI in conductive complex oxide materials and provide new opportunities to engineer and realize SOT-based devices in oxide electronics, potentially with new functionalities.

**Acknowledgement:** This work was supported by, and also made use of the Shared Facilities of, the Cornell Center for Materials Research with funding from the NSF MRSEC program (DMR-1719875). This work also made use of the Cornell NanoScale Facility/National Nanotechnology Coordinated Infrastructure which is supported by the NSF (ECCS-1542081). Z.W., H.P.N. and D.G.S. gratefully acknowledge the support from a GRO 'functional oxides' project from the Samsung Advanced Institute of Technology and support from the W.M. Keck Foundation. H.P. acknowledges support by the National Science Foundation (Platform for the Accelerated Realization, Analysis, and Discovery of Interface Materials (PARADIM)) under Cooperative Agreement No. DMR-1539918. C.S.C. and electron microscopy characterization were supported by the Department of Energy (DE-SC0002334).

**Figure captions**:

Figure 1 (a) Schematic showing the crystal structure of SrRuO₃ on SrTiO₃. (b) Low magnification ABF image exhibiting the well-ordered heterostructure of STO/SRO/Co. (c) HAADF and ABF images show a clearly delineated interface between the sputtered Co and the top of the SRO. (d) Resistivity measurements of SRO(10 nm) as a function of temperature. Inset: The derivative of resistivity $d\rho_{xx}/dT$ versus $T$. The kink around 150 K indicates the Curie temperature of the SRO.

Figure 2 (a) ABF image along the [010] zone axis of the SRO orthorhombic crystallographic axis showing small octahedral tilts in the SRO layer within the SRO(6)/Co(6) sample. (b) Similar ABF image of the SRO layer within the SRO(10)/Co(6) sample showing larger octahedral tilts compared to Fig. 2a. (c) The average projected tilt angle for each layer of the SRO(6)/Co(6) and SRO(10)/Co(6) samples determined from the ABF images. The colored diagrams within (a) and (b) depict the tilted octahedra overlaid on the ABF images.

Figure 3 (a) Schematic of the SRO/Co spin torque geometry. (b) The SHE-induced anti-damping spin torque efficiency $\xi_D$ determined from the angular ST-FMR as a function of $T$. (c) The extracted effective spin Hall conductivity $\sigma_{SH}^{\text{eff}}$ of SRO as a function of $T$. Inset: the corresponding $T$-dependent charge conductivity of the 10 nm SRO layer measured in the same $T$ range. (d) The damping-like spin torque efficiency arising from the spin current component with perpendicular spin polarization $\hat{s}_z$, and (inset) the perpendicular effective field, plotted as a function of $T$.

Figure 1

(a)

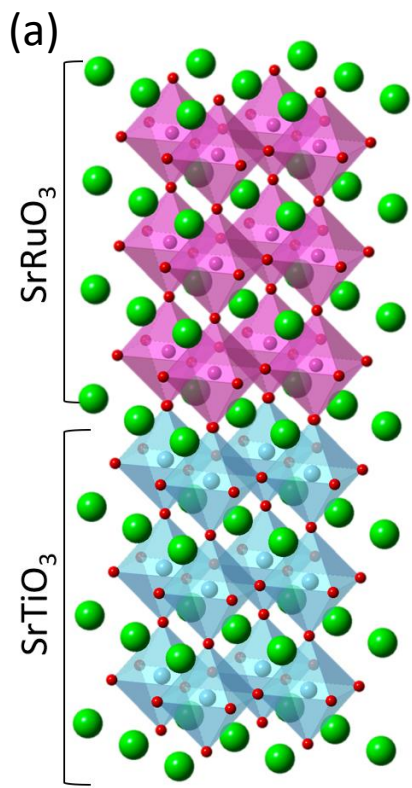

SrRuO₃

SrTiO₃

(b)

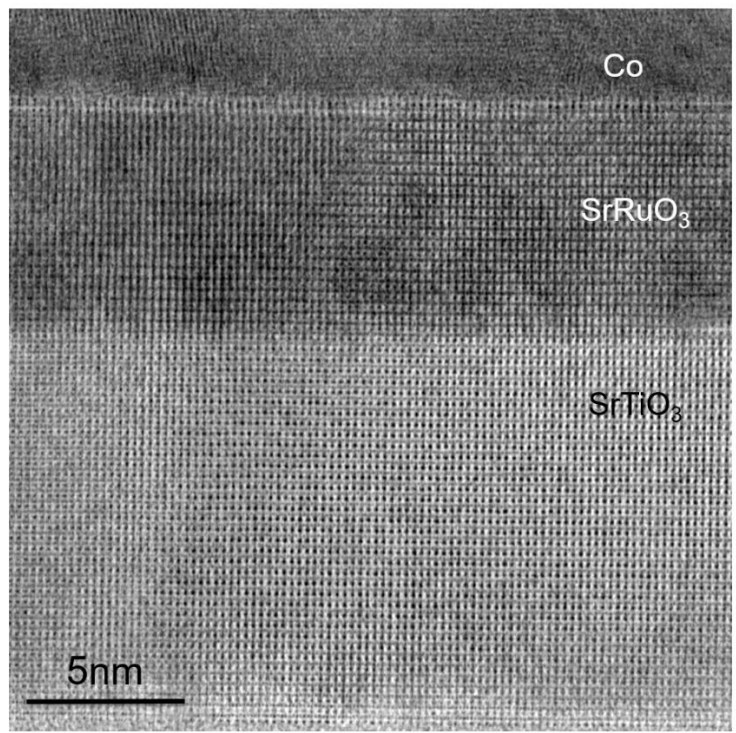

Co

SrRuO₃

SrTiO₃

5nm

(c)

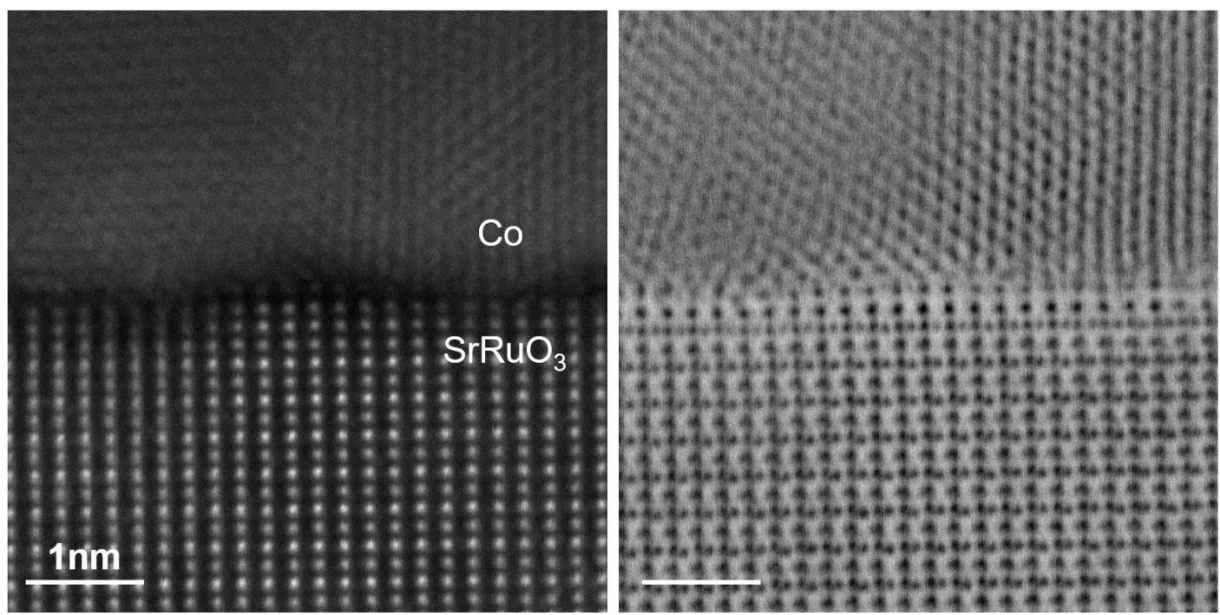

(d)

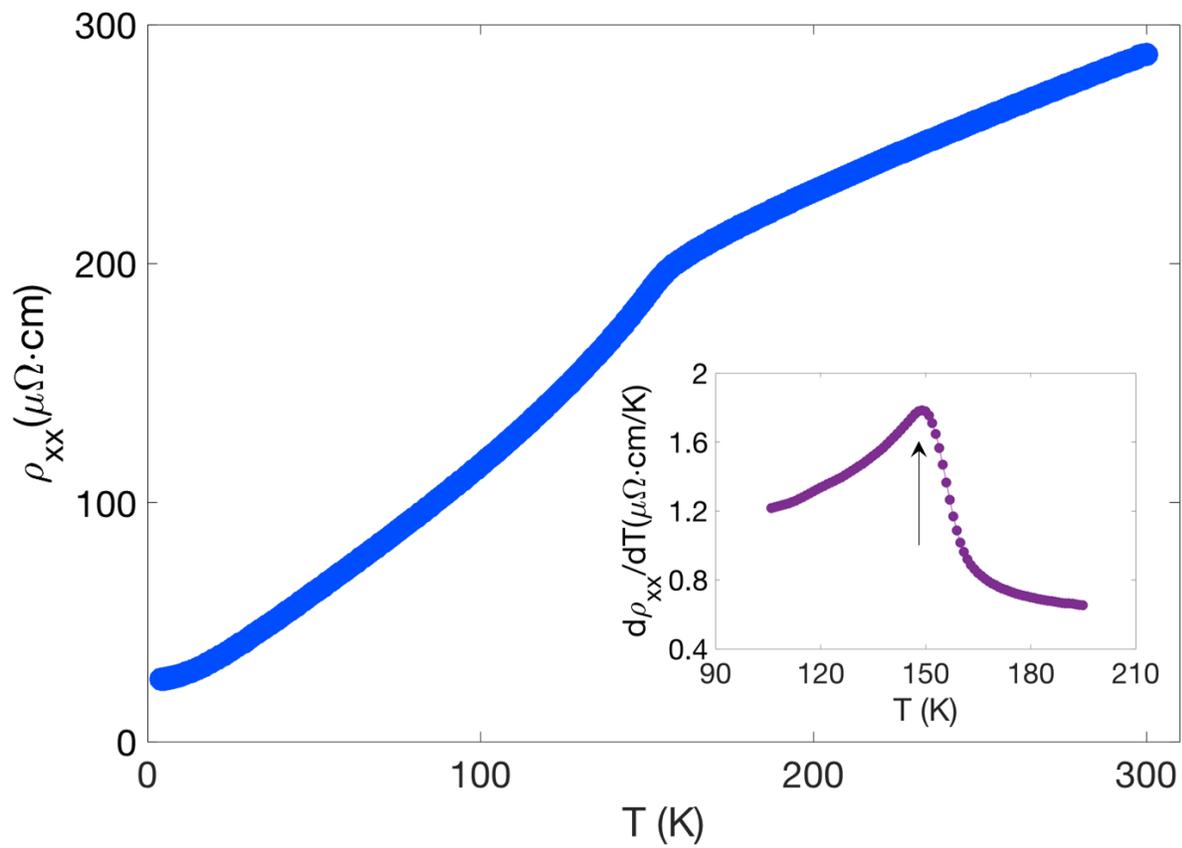

Figure 2

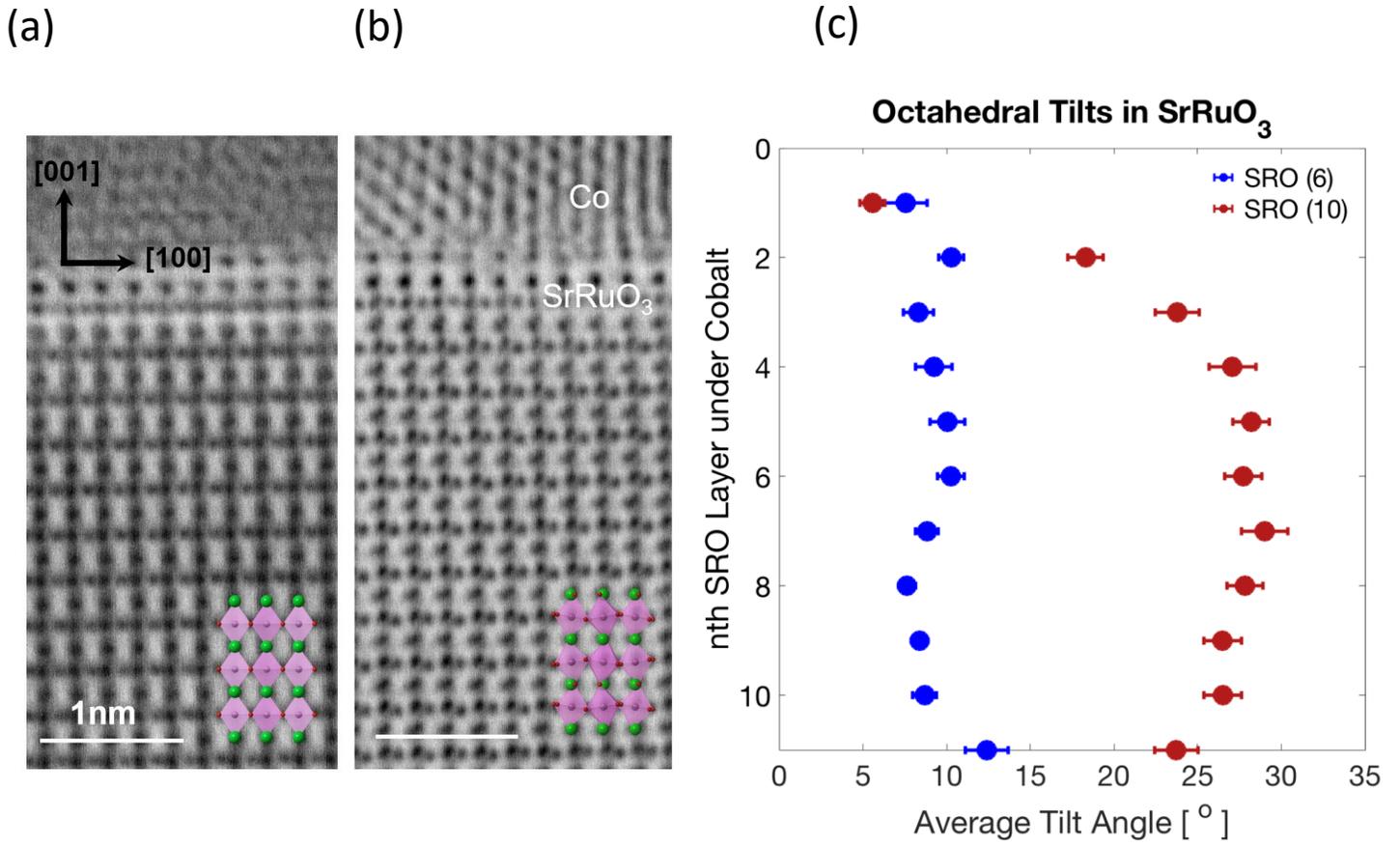

Figure 3

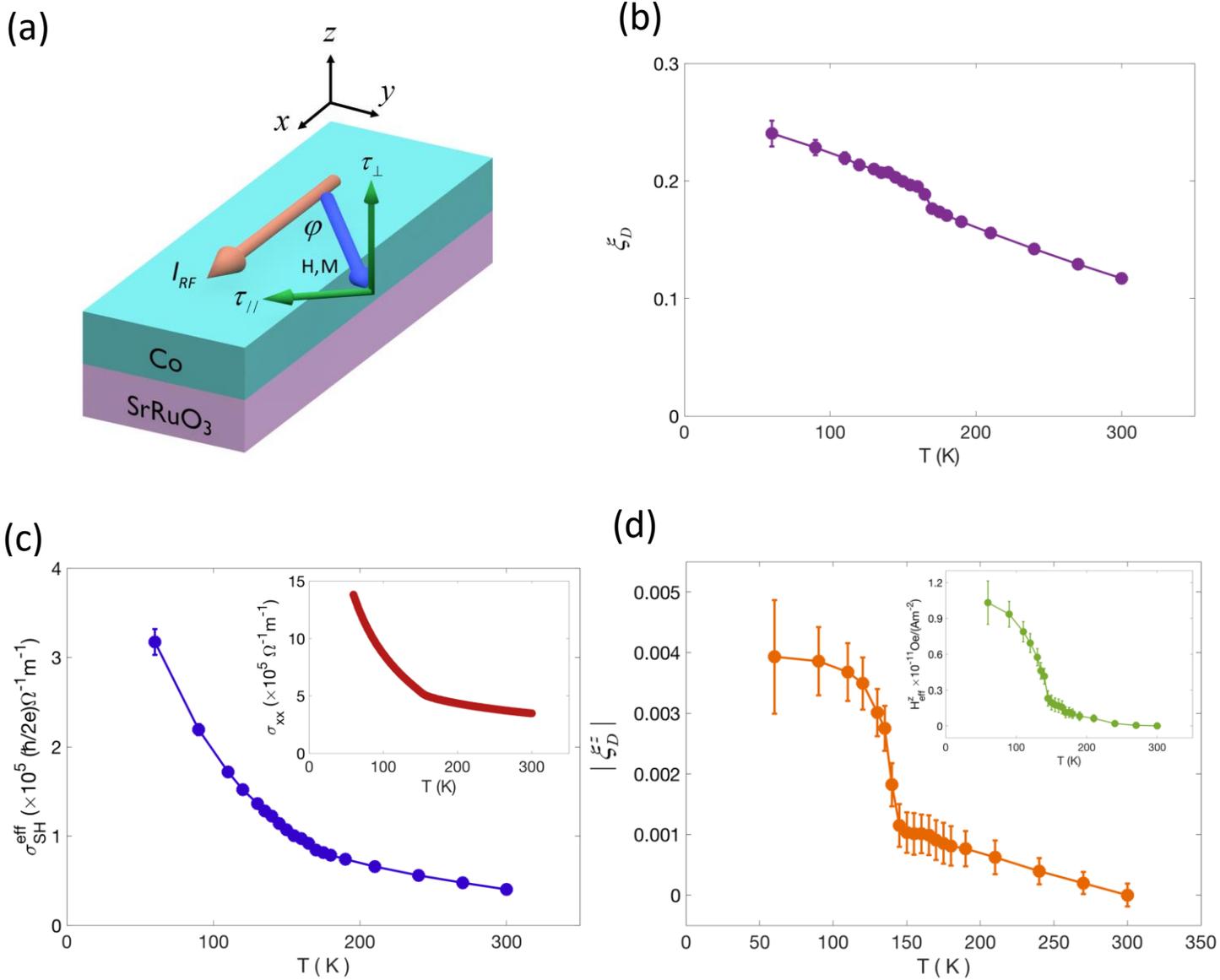

**Table 1**

| Sample | $\xi_D$ at 300K |
|---|---|
| SRO(10)/Co(6) | $0.10 \pm 0.02$ |
| SRO(6)/Co(6) | $0.05 \pm 0.01$ |
| SRO(4)/Co(6) | $0.05 \pm 0.01$ |
| SRO(10)/Co(4) | $0.10 \pm 0.01$ |
| SRO(6)/Co(4) | $0.06 \pm 0.01$ |
| SRO(4)/Co(4) | $0.05 \pm 0.01$ |

Table 1: The anti-damping spin torque efficiency $\xi_D$ determined from ST-FMR at 300 K for different SRO($d_{SRO}$)/Co($t_{Co}$) bilayers. The RF frequency range is between 12 GHz and 16 GHz.

**Exceptionally high, strongly temperature dependent, spin Hall conductivity of SrRuO₃**

Yongxi Ou[1,2*†], Zhe Wang[1†], Celesta S. Chang[1,2†], Hari P. Nair[3], Hanjong Paik[3,4], Neal Reynolds[2], D.C. Ralph[2,5], D. A. Muller[1,5], D. G. Schlom[3,5] and R. A. Buhrman[1*]

[1]School of Applied and Engineering Physics, Cornell University, Ithaca, New York 14853, USA,
[2]Department of Physics, Cornell University, Ithaca, New York 14853, USA,
[3]Department of Materials Science and Engineering, Cornell University, Ithaca, New York 14853, USA,
[4]Platform for the Accelerated Realization, Analysis, & Discovery of Interface Materials (PARADIM), Cornell University, Ithaca, New York 14853, USA,
[5]Kavli Institute at Cornell for Nanoscale Science, Ithaca, New York 14853, USA
*email: yo84@cornell.edu; rab8@cornell.edu.
†: These authors contributed equally to this work.

**Contents:**

**S1. STO/SRO/Co/Al thin film growth, fabrication and measurements**

**S2. X-ray diffraction characterization and High-resolution, aberration-corrected Transmission Electron Microscopy**

**S3. The dependence of SOTs on the magnetization direction and the direction of polarization of the incident spin current**

**S4. Definition and evaluation of the torque efficiencies from a transverse (out of plane) spin current $j_{sz}$ with $\hat{S}_z$ spin polarization**

**S5. Discussion regarding the possibility of a *T*-dependent interfacial spin transparency $T_{\text{int}}$.**

**S6. *T*-dependent demagnetization field of the Co layer**

**S7. Comparison of *T*-dependent spin torques in two different microstrips patterned from the same SRO(10)/Co(6) sample**

**S1. STO/SRO/Co/Al thin film growth, fabrication and measurements**

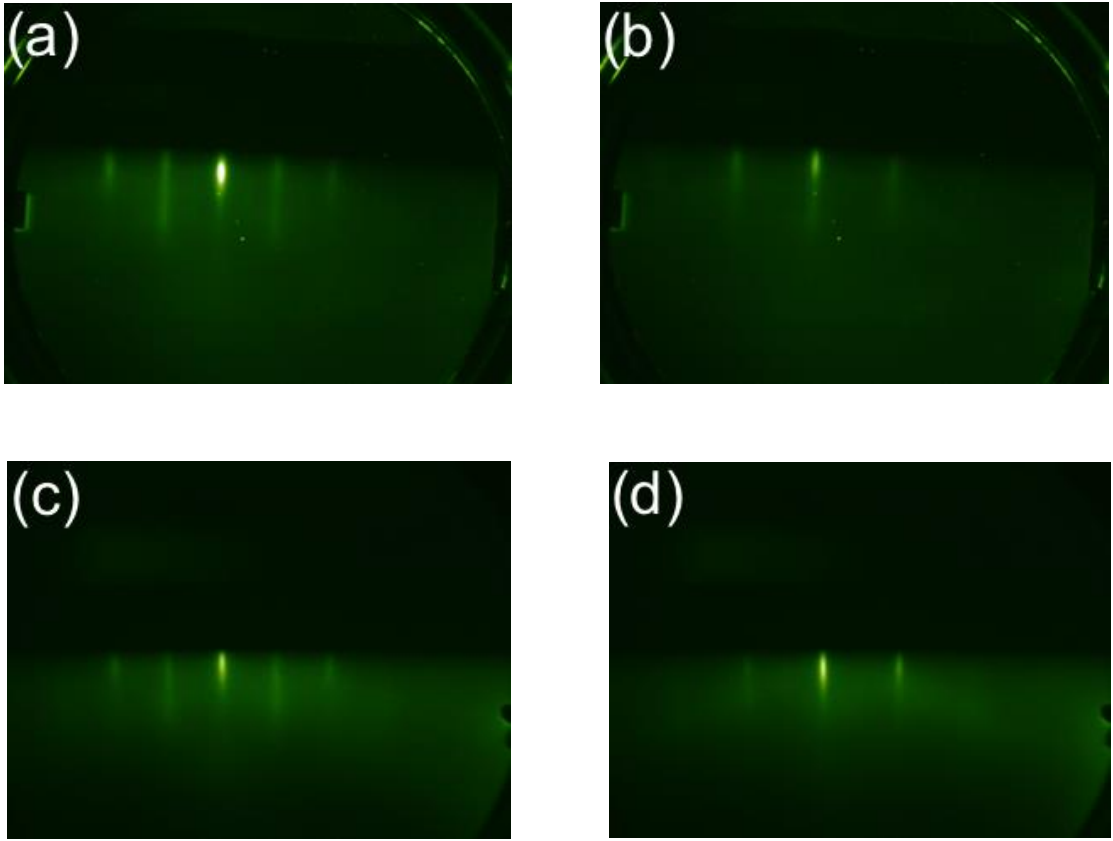

Figure S1: RHEED patterns of a 15 nm thick SrTiO$_3$ film along (a) the [100] and (b) the [110] azimuth of SrTiO$_3$. RHEED patterns of a 10 nm thick SrRuO$_3$ along (c) the [100]$_p$ and (d) the [110]$_p$ azimuth of SrRuO$_3$.

The SrTiO$_3$, SrRuO$_3$, Co and Al capping layer were all grown in a Veeco Gen10 dual-chamber MBE system on 2″ commercial silicon wafers (*p*-type, boron doped, and resistivity = >10Ω•cm ). To be specific, the 15 nm thick SrTiO$_3$ buffer layer was grown in the first chamber with a base pressure in the upper $10^{-9}$ Torr range. The wafer was then transferred in vacuum to a second growth chamber with a base pressure in the lower $10^{-8}$ Torr range. Following the growth of the SrRuO$_3$ in the second chamber the stack was transferred back to the first growth chamber for deposition of the Co layer and the Al layer. Both growth chambers are equipped with *in-situ* reflection high-energy electron diffraction (RHEED) systems for monitoring the film growth. Substrate temperature was monitored either by a thermocouple for temperatures below 500 °C or an optical pyrometer with a measurement wavelength of 980 nm for temperatures above 500 °C. Prior to film growth, the silicon substrate was cleaned *ex situ* in an ultraviolet ozone cleaner for 20 min to remove organic contaminants from the surface of the substrate. Molecular beams of strontium, titanium and ruthenium were generated from elemental sources using a conventional low-temperature effusion cell, a Ti-Ball$^{TM}$ [1], and an electron-beam evaporator, respectively.

The SrTiO$_3$ layer in our samples was grown in the first growth chamber by the epitaxy-by-periodic-annealing method [2] for its first 2 nm (5 unit cells) and then with a high temperature codeposition (strontium, titanium and oxygen all supplied simultaneously) growth step at 580 °C under $5 \sim 8 \times 10^{-8}$ Torr O$_2$ to achieve a total SrTiO$_3$ film thickness of 15 nm. The growth rate of the SrTiO3 film was $\sim 70$ s per unit cell. Growth of the SrTiO$_3$ layer on silicon is described in detail elsewhere [3,4].

The SrRuO$_3$ film was grown under adsorption-controlled growth conditions [5] after transfer to the second growth chamber under vacuum. Unlike the growth of SrTiO$_3$, which needs careful calibration to provide 1:1 matched fluxes of strontium and titanium [6] to yield a stoichiometric SrTiO$_3$ film [7], the stoichiometry of the SrRuO$_3$ film grown by adsorption-controlled growth is ensured by providing an excess ruthenium flux to the growing film and exploiting thermodynamics to precisely desorb the excess ruthenium in the form RuO$_x$(g). We grew the SrRuO$_3$ film at a substrate temperature of 660-700 °C (measured using the optical pyrometer) and an oxidant (a mixture of ~10% O$_3$ + 90% O$_2$) background pressure of $1 \times 10^{-6}$ Torr. More details of the growth of the SrRuO$_3$ film can be found elsewhere [3,5].

The polycrystalline Co and Al layers were both deposited at room temperature in the first growth chamber under vacuum. The Al layer was fully oxidized into amorphous aluminum oxide when the heterostructure was taken out of vacuum. The molecular beams of cobalt and aluminum were both generated from medium-temperature effusion cells.

RHEED patterns of a 15 nm thick SrTiO$_3$ film on silicon along the [100] and [110] azimuths of SrTiO$_3$ are shown in Figs. S1(a) and (b), respectively. SRO has an orthorhombic GdFeO$_3$-type perovskite structure with lattice constants a = 5.53283 Å, b = 5.56296 Å and c = 7.84712 Å corresponding to the non-standard setting *Pbnm* of space group #62 [8]. The orthorhombic crystal structure of SrRuO$_3$ is derived from the ideal perovskite structure through a tilt of the RuO$_6$ octahedra about the b-axis and in-phase rotation about the c-axis. RHEED patterns of a 10 nm thick SrRuO$_3$ film on top of 15 nm thick SrTiO$_3$ on silicon along the [100]$_p$ (where the subscript *p* denotes pseudocubic indices) and the [110]$_p$ azimuths of SrRuO$_3$ are shown in Figs. S1(c) and (d), respectively. No phases other than the perovskite phase were detected during or after the growth of the SrTiO$_3$ and SrRuO$_3$ films. This combined with the streaky RHEED patterns indicates that both the SrTiO$_3$ and SrRuO$_3$ films are single-phase and epitaxial.

After the deposition, the $SrTiO_3/SrRuO_3/Co/Al$ thin films were then patterned into $20\times10\,\mu m$ microstrips via standard photolithography and ion mill etching processes. During the ion mill etching, the elemental-sensitive etching process was monitored via endpoint detection and stopped right before etching into the $SrTiO_3$ to avoid creating conducting defects in the $SrTiO_3$ layer. Resistivity measurements indicate the remaining $SrTiO_3$ buffer remained insulating while the total resistance of the $SrRuO_3/Co$ bilayers was consistent with the parallel resistance of the $SrRuO_3$ and Co. After the ion etching, Ti/Pt contact pads were deposited in a ground-signal-ground geometry [9].

The resistivity ($\rho$) vs. temperature ($T$) of a $Si/SrTiO_3(15)/SrRuO_3(10)$ sample was measured in a standard four-probe van der Pauw geometry with wire-bonded contacts made using aluminum wire in a Quantum Design physical property measurement system (PPMS).

For temperature dependent ST-FMR measurements, the SRO/Co bilayer samples were wire-bonded to a RF waveguide for RF current input, mounted inside a cryostat cooled by liquid helium. At each measured temperature, an RF current (13.5GHz) generated by a RF signal generator was applied to the $SrRuO_3/Co$ bilayer through a bias tee. An electromagnet was mounted on a rotational stage to provide a magnetic field that could be swept from 0.2 T to 0 T at a given in-plane angle relative to the RF current direction. The ST-FMR resonance signal was detected as a dc mixing voltage by a lock-in amplifier.

**S2. X-ray diffraction characterization and High-resolution, aberration-corrected Transmission Electron Microscopy**

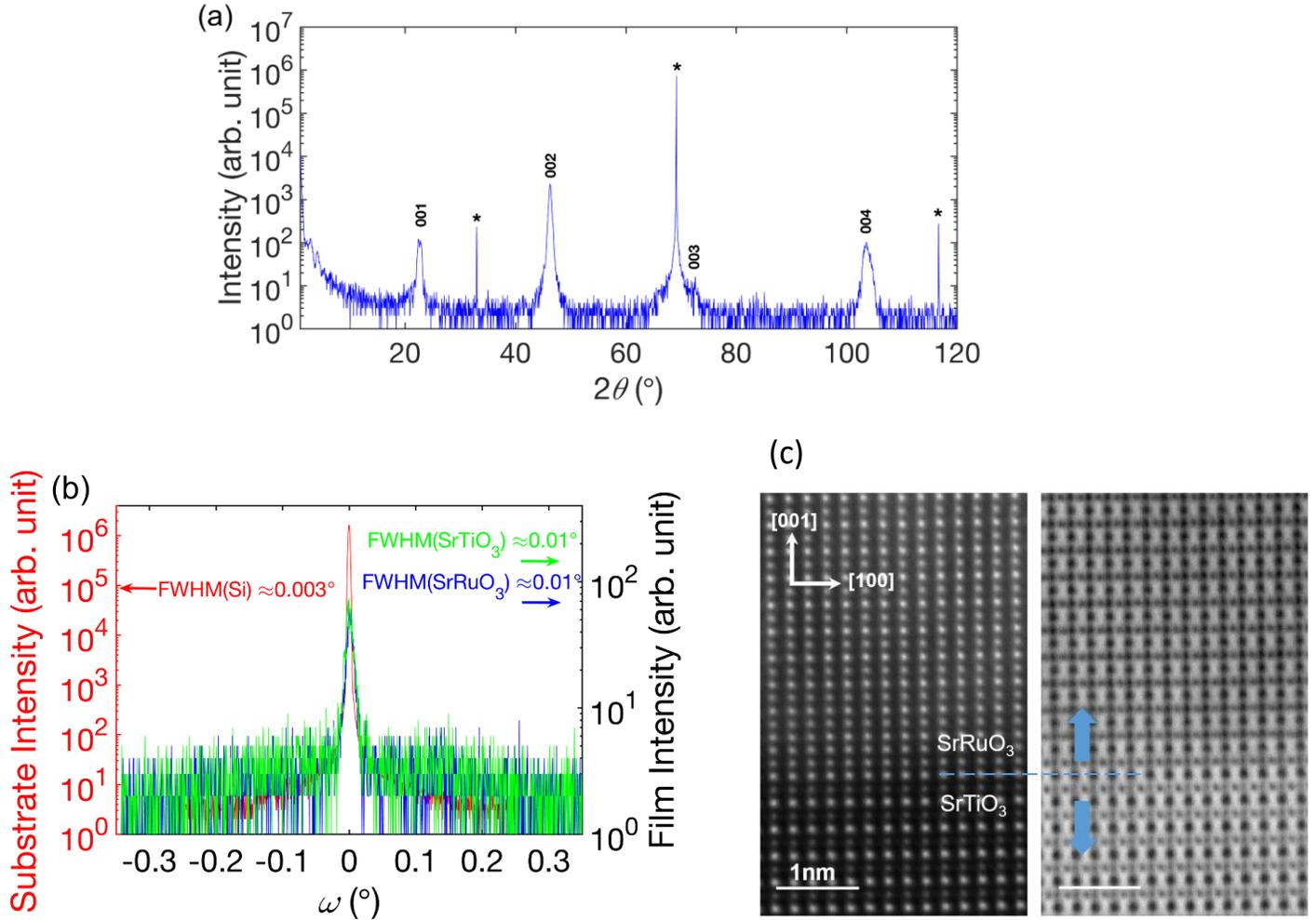

Figure S2: (a) XRD $\theta$-$2\theta$ scan of a Si/ SrTiO$_3$(15)/ SrRuO$_3$ (10)/Co(6)/ Al$_2$O$_3$ (1). (b) XRD rocking curves of a Si/ SrTiO$_3$(15)/ SrRuO$_3$ (10)/Co(2)/ Al$_2$O$_3$ (1) sample. (c) Simultaneously-acquired HAADF image (left) and ABF-image (right) show a sharp interface between SRO and STO.

The structural quality of the STO/SRO heterostructures were characterized via *ex situ* x-ray diffraction (XRD) with a PANalytical X'pert system utilizing Cu $K\alpha_1$ radiation. As is shown in Fig. S2(a), a XRD $\theta$-$2\theta$ scan of the 1 nm thick $Al_2O_3$ on 6 nm thick Co on 10 nm thick $SrRuO_3$ on 15 nm thick $SrTiO_3$ on silicon shows that the $SrTiO_3$ and $SrRuO_3$ films are epitaxial, with only 00$l$ reflections present. Rocking curve measurements on a similar sample show that both the $SrTiO_3$ and $SrRuO_3$ films are of high crystalline perfection, with a full width at half maximum of both the $SrTiO_3$ 001 and the $SrRuO_3$ 001$_p$ peaks being ~0.01°, as is shown in Fig. S2(b). The rocking curve of the Si 004 peak is overlaid for comparison.

Cross-sectional TEM specimens were prepared using an FEI Strata 400 Focused Ion Beam (FIB) with a final milling step of 5 keV to reduce surface damage. High-resolution ABF and HAADF-STEM images were acquired on an aberration corrected 300-keV Themis Titan. Figure S2(c) shows high magnification high-angle annular dark-field (HAADF) and ABF STEM images of the STO/SRO interface.

The projected octahedral tilts were quantified layer by layer from ABF-STEM images. Stacks of images were acquired and averaged to reduce scan noise. The atomic positions of ruthenium and oxygen were determined using an open matlab code StatSTEM written by A. De Backer [10].

**S3. The dependence of SOTs on the magnetization direction and the direction of polarization of the incident spin current**

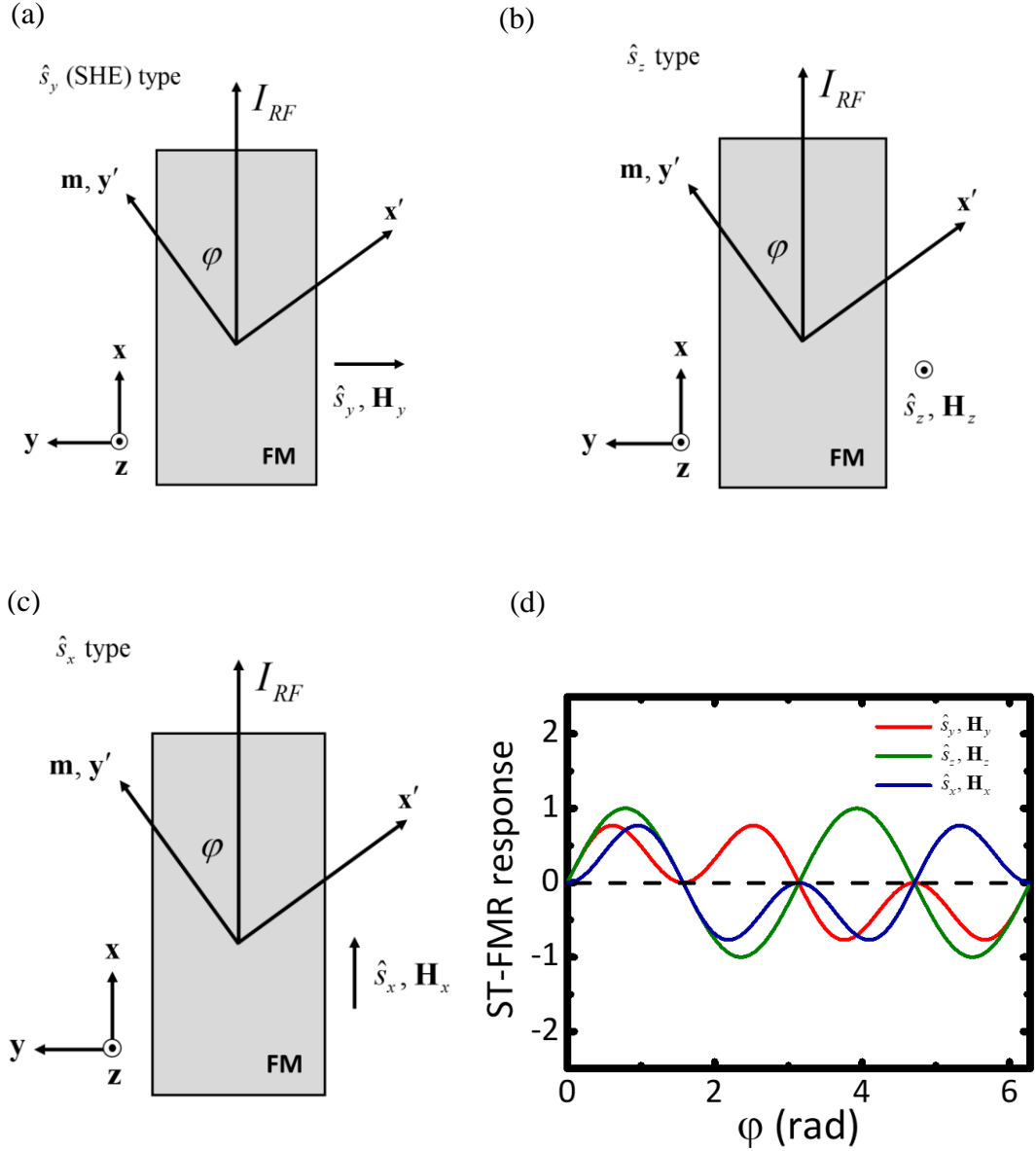

Figure S3: (a-c) Schematic illustration showing three possible polarization directions of an incident spin current, $\hat{s}_z, \hat{s}_y, \hat{s}_x$, and the associated current-induced effective fields: (a) $\hat{s}_y$ (SHE), (b) $\hat{s}_z$ and (c) $\hat{s}_x$. (d) The functional forms of the angular dependence of the ST-FMR response.

The current-induced SOTs exerted on a ferromagnet (FM) layer are most commonly the result of an injected spin current from an adjacent material that possesses a strong SOI, e.g. the SHE. The dynamic response of the FM's magnetization as the result of this spin current injection follows the generalized Landau–Lifshitz–Gilbert-Slonczewski (LLGS) equation:

$$\frac{d\mathbf{m}}{dt} = -\gamma\mathbf{m}\times\mathbf{H}_{mag} + \alpha\mathbf{m}\times\frac{d\mathbf{m}}{dt} + \tau_D\mathbf{m}\times\left(\hat{s}\times\mathbf{m}\right) + \tau_F\mathbf{m}\times\mathbf{H}_F \qquad (S3.1)$$

where $\mathbf{m}$ is the unit vector of the magnetization, $\mathbf{H}_{mag}$ is the static, externally applied magnetic field, $\gamma$ is the gyromagnetic ratio, and $\alpha$ is the Gilbert damping constant of the FM. The third and fourth terms on the right side of Eqn. S3.1 correspond to the damping-like (DL) and field-like (FL) torques, where $\hat{s}$ and $\hat{\mathbf{H}}_F$ are the unit vectors of the spin polarization and current-induced field (the latter including both the Oersted field from the RF current in the non-magnetic layer, and any "field-like" effective field generated by the incident spin current), while $\tau_D$ and $\tau_F$ denote the strength of these two terms. For a given FM system with in-plane magnetization $\mathbf{m}$ (Co in our case), the DL and FL torques generated by an in-plane current $j_e\hat{x}$ can be either in-plane (IP, //) or out-of-plane (OP, $\perp$), depending on the directions of $\hat{s}$ and $\mathbf{H}_F$.

(a) SHE induced SOTs - $\hat{s}_y$ spin current polarization

As is illustrated in Fig. S3(a), the spin current generated by the SHE in a normal metal exhibits an in-plane spin polarization $\hat{s}_y$ transverse to the current direction. In this case, the angular dependence of the resultant DL torque is $\tau_D^0\mathbf{m}\times(\hat{s}_y\times\mathbf{m}) \propto \tau_D^0\cos\varphi\mathbf{x}'$, that is, an IP torque. The Oersted field from the RF current is also transverse to the current direction, resulting in a FL term $\tau_F^0 \mathbf{m} \times \mathbf{H}_y \propto \tau_F^0 \cos\varphi \mathbf{z}'$, an OP torque. If there were also a field-like torque on the FM resulting from the spin current, it also would create an OP torque that would either add to, or subtract from the Oersted field torque, with a strength relative to the Oersted field torque that would vary inversely with the thickness of the FM. However, as mentioned in the main text in our SRO/Co samples any such field-like torque from the SHE appears to be negligible compared to that from the Oersted field.

(b) SOTs from a spin current component with $\hat{s}_z$ polarization and the related effective field $H_z$

As shown in Fig. S3(b), if the polarization and any related spin-current induced effective field has an OP component, namely $\hat{s}_z$ and $\mathbf{H}_z$, then the results are DL and FL torques with symmetry that is readily distinguishable from that of the torques arising from the SHE, case (a) above. In this latter case, the OP spin polarization generates a DL torque $\tau_D^z \mathbf{m} \times (\hat{s}_z \times \mathbf{m}) \propto \tau_D^z \mathbf{z}'$, which is an OP torque, while the OP effective field generates a FL torque $\tau_F^z \mathbf{m} \times \mathbf{H}_z \propto \tau_F^z \mathbf{x}'$, which is an IP torque. Note that here, both torques are independent of the in-plane magnetization direction of the FM.

(c) SOTs from a spin current component with $\hat{s}_x$ polarization and the related effective field $H_x$

In addition to the two cases above, the polarization of the spin current, and any related spin-current induced effective field could, in principle, also include a component parallel to the current direction ($\hat{x}$) (Fig. S3(c)), *e.g.* as part of a Dresselhaus-type SOT [11,12]. If present, that spin current component will generate a DL torque $\tau_D^x \mathbf{m} \times (\hat{s}_x \times \mathbf{m}) \propto \tau_D^x \sin\varphi \mathbf{x}'$, which is an IP torque, while the effective field will generate a FL torque $\tau_F^x \mathbf{m} \times \mathbf{H}_x \propto \tau_F^x \sin\varphi \mathbf{z}'$, which is an OP torque.

In a ST-FMR measurement any IP torque generates a resonant FMR signal that is symmetric about the resonant bias field, and any OP torque generates a signal that is antisymmetric about the resonant field. Since the FMR resonance is detected by the anisotropic magnetoresistance (AMR) [13] of the FM (Co in our case) , both the symmetric and antisymmetric components of the ST-FMR voltage signal will be modulated by $\sin 2\varphi$. Thus in general the two components of the ST-FMR signal will exhibit the following angular dependence:

$$V_s = V_s^y \cos\varphi \sin 2\varphi + V_s^z \sin 2\varphi + V_s^x \sin\varphi \sin 2\varphi \qquad (S3.2)$$

$$V_A = V_A^y \cos\varphi \sin 2\varphi + V_A^z \sin 2\varphi + V_A^x \sin\varphi \sin 2\varphi \qquad (S3.3)$$

the same as Eq.(2) in the main text, with the magnitude of the spin current components in the three possible orthogonal directions, and the related effective field components, determining the amplitude of each term in Eqn. S3.2/Eqn. S3.3. Table S3 summarize the symmetries for the SOTs and their induced ST-FMR signals for the above three cases. Fig. S3d plots the functional forms for the three voltage components as a function of magnetization angle, assuming for the illustration that all three components have the same amplitude.

**Table S3**

| | Torque type | IP or OP | Torque Symmetry | ST-FMR signal type | ST-FMT signal symmetry |
|---|---|---|---|---|---|
| **(a)** | DL | IP | $\cos\varphi$ | symmetric | $\cos\varphi\sin2\varphi$ |
| $\hat{s}_y$ **(SHE)** | FL | OP | $\cos\varphi$ | antisymmetric | $\cos\varphi\sin2\varphi$ |
| **(b)** | DL | OP | 1 | antisymmetric | $\sin2\varphi$ |
| $\hat{s}_z$ | FL | IP | 1 | symmetric | $\sin2\varphi$ |
| **(c)** | DL | IP | $\sin\varphi$ | symmetric | $\sin\varphi\sin2\varphi$ |
| $\hat{s}_x$ | FL | OP | $\sin\varphi$ | antisymmetric | $\sin\varphi\sin2\varphi$ |

**S4. Definition and evaluation of the torque efficiencies from a transverse (out of plane) spin current $j_{sz}$ with $\hat{s}_z$ spin polarization**

Due to the different nature of the torque symmetry for the spin torques from a spin current component with $\hat{s}_z$ polarization, described in section S3, it is necessary to carefully define the quantity of the corresponding torque efficiencies.

In the ST-FMR measurement, a general in-plane torque results in a symmetric component that is scaled by the magnitude of this in-plane torque, $\tau_{//}$, while a general out-of-plane torque corresponds to a antisymmetric component that is scaled to the magnitude of this out-of-plane torque modified by a pre-factor that depends on the demagnetization field $M_{\text{eff}}$ and the resonance field $H_{\text{res}}$, $t_\wedge \sqrt{1 + M_{\text{eff}} / H_{\text{res}}}$.

As summarized in Table S3, in the SHE scenario, the DL torque is an *in-plane* torque that results in the symmetric component $V_S^0$ (i.e. $V_S^y$) while the Oersted field from the RF current generates an *out-of-plane* FL torque that results in the antisymmetric component $V_A^0$ (i.e. $V_A^y$). The ratio of these two components gives [13]:

$$\frac{V_S^0}{V_A^0} = \frac{\tau_{//}^0}{\tau_\perp^0 \sqrt{1 + M_{\text{eff}} / H_{\text{res}}}} \qquad (S4.1)$$

Notice that in this case $\tau_{//}^y = \gamma(\hbar / 2e\mu_0 M_s t)J_s^0$ where $J_s^0$ is the spin current density generated by the SHE (with spin polarization directed in-plane), and $t_\wedge^y = g H_{RF} = g J_c d / 2$ where $J_c$ is the charge current density. Therefore, Eq.(S4.1) can be rewritten as:

$$\frac{V_S^y}{V_A^y} = \frac{\tau_{//}^y}{\tau_{\perp}^y \sqrt{1 + M_{\text{eff}} / H_{\text{res}}}} = \frac{(\hbar / 2e\mu_0 M_s t) J_s^0}{J_c d / 2\sqrt{1 + M_{\text{eff}} / H_{\text{res}}}} \qquad (S4.2)$$

From Eq.(S4.2) we can define and calculate the DL spin torque efficiency, $\xi_D$, as the ratio of $J_s^0$ to $J_c$ [13]:

$$\xi_D \equiv \frac{J_s^0}{J_c} = \frac{V_S^y}{V_A^y} \frac{e\mu_0 M_s t d}{\hbar} \sqrt{1 + M_{\text{eff}} / H_{\text{res}}} \qquad (S4.3)$$

This is the Eq.(1) in the main text.

For the case of a spin current component with $\hat{s}_z$ polarization, one needs to be careful since the symmetry of these torques are different from the SHE-type torques. As listed in Table S3, the $s_z$ DL torque is an out-of-plane torque that corresponds to an antisymmetric ST-FMR signal, $V_A^z \propto \tau_{\perp}^z \sqrt{1 + M_{\text{eff}} / H_{\text{res}}}$. The $\hat{s}_z$ FL torque is an in-plane torque that corresponds to a symmetric ST-FMR signal, $V_S^z \propto \tau_{//}^z$. A reasonable way to compare the different torques way is to the compare these torques (*i.e.* their ST-FMR signal amplitudes) to the signal from the RF current (i.e. Oersted field, $V_A^y$), as in the SHE case summarized above:

$$\frac{V_A^z}{V_A^0} = \frac{\tau_{\perp}^z \sqrt{1 + M_{\text{eff}} / H_{\text{res}}}}{\tau_{\perp}^y \sqrt{1 + M_{\text{eff}} / H_{\text{res}}}} = \frac{\tau_{\perp}^z}{\tau_{\perp}^y} = \frac{\gamma(\hbar / 2e\mu_0 M_s t) J_s^z}{\gamma J_c d / 2} \qquad (S4.4)$$

$$\frac{V_S^z}{V_A^y} = \frac{t_{//}^z}{t_{\wedge}^y \sqrt{1 + M_{\text{eff}} / H_{\text{res}}}} = \frac{g H_{\text{eff}}^z}{g H_{RF} \sqrt{1 + M_{\text{eff}} / H_{\text{res}}}} \qquad (S4.5)$$

Note that we have parameterized $\tau_\perp^z$ and $\tau_{//}^z$ by the spin current component that has perpendicular spin polarization (i.e. $J_s^z$) and the perpendicular effective field (i.e. $H_{\text{eff}}^z$) respectively in the equations above. With these relations, we can evaluate the following damping-like spin torque efficiency and effective field:

$$\xi_D^z \equiv \frac{J_s^z}{J_c} = \frac{V_A^z}{V_A^0} \frac{e\mu_0 M_s t d}{\hbar} \qquad (S4.6)$$

$$H_{\text{eff}}^z \equiv H_{RF} \frac{V_S^z}{V_A^0} \sqrt{1 + M_{\text{eff}} / H_{\text{res}}} \qquad (S4.7)$$

The physical meaning of the two quantities in Eq.(S4.6) and Eq.(S4.7) are clear: $\xi_D^z$ is similar to $\chi_D$ in the SHE scenario and describes the efficiency at which the charge current generates a spin current with perpendicular spin polarization that results in an out-of-plane damping-like torque on the FM; $H_{\text{eff}}^z$ describes the strength of the perpendicular effective field. We plot $\xi_D^z(H_{\text{eff}}^z)$ as a function of $T$ in Fig. 3d (and its inset) in the main text.

Similarly, we can define the DL spin torque efficiency and effective field of the spin torques from a spin current with $\hat{s}_x$ polarization from the ST-FMR signals with the help of Table S3:

$$\xi_D^x = \frac{J_s^x}{J_c} = \frac{V_S^x}{V_A^0} \frac{e\mu_0 M_s t d}{\hbar} \sqrt{1 + M_{\text{eff}} / H_{\text{res}}} \qquad (S4.8)$$

$$H_{\text{eff}}^x \equiv H_{RF} \frac{V_A^x}{V_A^0} \qquad (S4.9)$$

As mentioned in the main text, there is not a clearly discernable spin current component with such a polarization in our ST-FMR measurement of the SRO/Co samples.

**S5. Discussion regarding the possibility of a *T*-dependent interfacial spin transparency $T_{int}$.**

As discussed in the main text and plotted in Fig. 3b, we have measured a damping-like spin torque efficiency $\chi_{DL}(T)$ in our SRO/Co material that is strongly *T*-dependent, with $\chi_{DL}(T)$ becoming higher as the electrical conductivity $S_{xx}^{SRO}(T)$ of the SRO also increases with decreasing *T*. Usually the spin Hall conductivity $S_{SH}$ of a material is expected to temperature independent, whether it arises from extrinsic, impurity scattering effects, or intrinsically from the Berry curvature of the band structure. This would result in $\xi_{DL} \equiv T_{int}\theta_{SH}(T) \equiv T_{int}(\hbar/2e)\sigma_{SH}/\sigma_{xx}(T)$ either being constant with *T, i.e.* with changes in $S_{xx}(T)$, for extrinsic skew scattering, or decreasing with decreasing *T* (increasing $S_{xx}(T)$) if side jump scattering or the intrinsic Berry curvature mechanism dominates. Thus, it appears that either there is a quite strong temperature variation to the SRO $S_{SH}$, or $T_{int}$ the interfacial spin transparency must be strongly temperature dependent. We consider the latter quite unlikely.

If $T_{int}$ is dominated by simple interfacial back-flow, then according to the standard drift-diffusion analysis it can be written as [14]:

$$T_{int}(T) = \frac{2G^{\uparrow\downarrow}(T)}{2G^{\uparrow\downarrow}(T) + \dfrac{\sigma_{xx}(T)}{\lambda_s(T)}} \quad \text{(S5.1)}$$

where $G^{\uparrow\downarrow}(T)$ is the interfacial spin mixing conductance [15], and $l_s$ is the spin diffusion length in the spin Hall material. The term $S_{xx}(T)/l_s(T)$ is usually referred to as the "spin conductance" $G_s(T)$.

$G^{\uparrow\downarrow}$ is closely related to the number of single electron channels bridging the interface [16], the Sharvin conductance, and thus is independent of temperature for typical metals. While SRO with its lower density of states and "bad metal" properties at room temperature is not necessarily a typical metal, there is no indication from the extensive studies that have been conducted on the transport properties of SRO to suggest that the overall density of states is strongly $T$ dependent as would be required to explain our experimental results for $\chi_{DL}(T)$ ($S_{xx}^{SRO}(T)$).

The spin conductance $G_s(T)$ depends on the dominant mechanism by which spins relax in the spin Hall material. In "dirty" metals with a short elastic mean free path $l_e$ and a strong spin-orbit interaction, the Elliot-Yafet spin mechanism should dominant and the spin diffusion length should scale with $l_e$. In that case $G_s$ is independent of $l_e$, and thus $G_s$ for SRO should also be independent of $T$. We conclude that it is quite unlikely that $T_{int}$ has a significant $T$-dependence. Thus while a strong increase in $S_{SH}(T)$ with decreasing $T$ is something of a surprising result, it seems by far the most probable explanation for our experimental results.

## S6. *T*-dependent demagnetization field of the Co layer

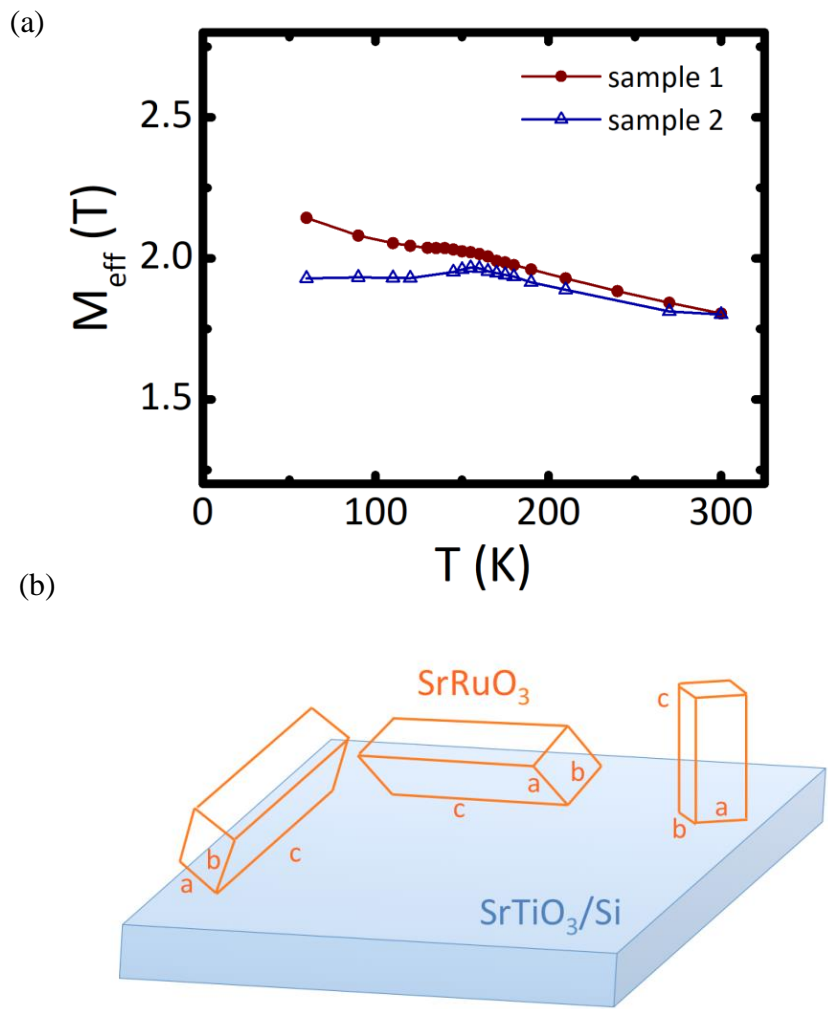

(a)

(b)

Figure S6: (a) The demagnetization field $M_{eff}$ of Co as a function of T for two different SRO(10)/Co(6) microstrip samples. (b) Schematic drawing of the twin configuration that shows the three possible orientations of the twinned SrRuO₃ film when grown on SrTiO₃ on silicon.

Below its Curie temperature (~ 150 K in our samples), SRO becomes ferromagnetic. As discussed in the main text, since our SRO films have twinned domains, some parts in the material will have the easy axis (b-axis) oriented at 45º to the plane of the film, and each microstrip can be expected to contain a variable amount of such material given the typical, ~ µm, scale of the twin domains. At the SRO/Co interface, this strong out-of-plane magnetic anisotropy in part of the SRO layer will result in a out-of-plane field being exerted on the Co layer, albeit not a spatially uniform one. As a result, one can expect that the demagnetization field of Co when the SRO/Co microstrip is below the SRO $T_c$ will be affected. Fig. S6, we show the demagnetization field $M_{eff}$ of Co for two SRO(10)/Co(6) microstrip samples as a function of $T$, as determined from the Kittel formula for the ferromagnetic resonance, $f = \frac{\mu_0 \gamma}{2\pi} \sqrt{H_{res}(H_{res} + M_{eff})}$. Above $T_c$, $M_{eff}$ increases gradually with decreasing $T$ as would be expected from the normal Bloch law temperature dependence of the saturation magnetization $M_s$ of Co. As can be clearly seen in Fig. S6(a), when $T$ decreases below $T_c$, $M_{eff}(T)$ immediately begins to deviate significantly from the trend seen above $T_c$, with, depending on the sample, the $M_{eff}(T)$ increase becoming slower or stopping, which we take as being indicative of OP magnetic domains in the SRO. The different $T$-dependent $M_{eff}$ behaviors between the two samples is consistent with a sample-dependent variation in the twinned domain configuration in different microstrips with possible orientations as illustrated in Fig. S6(b).

**S7. Comparison of *T*-dependent spin torques in two different microstrips patterned from the same SRO(10)/Co(6) sample**

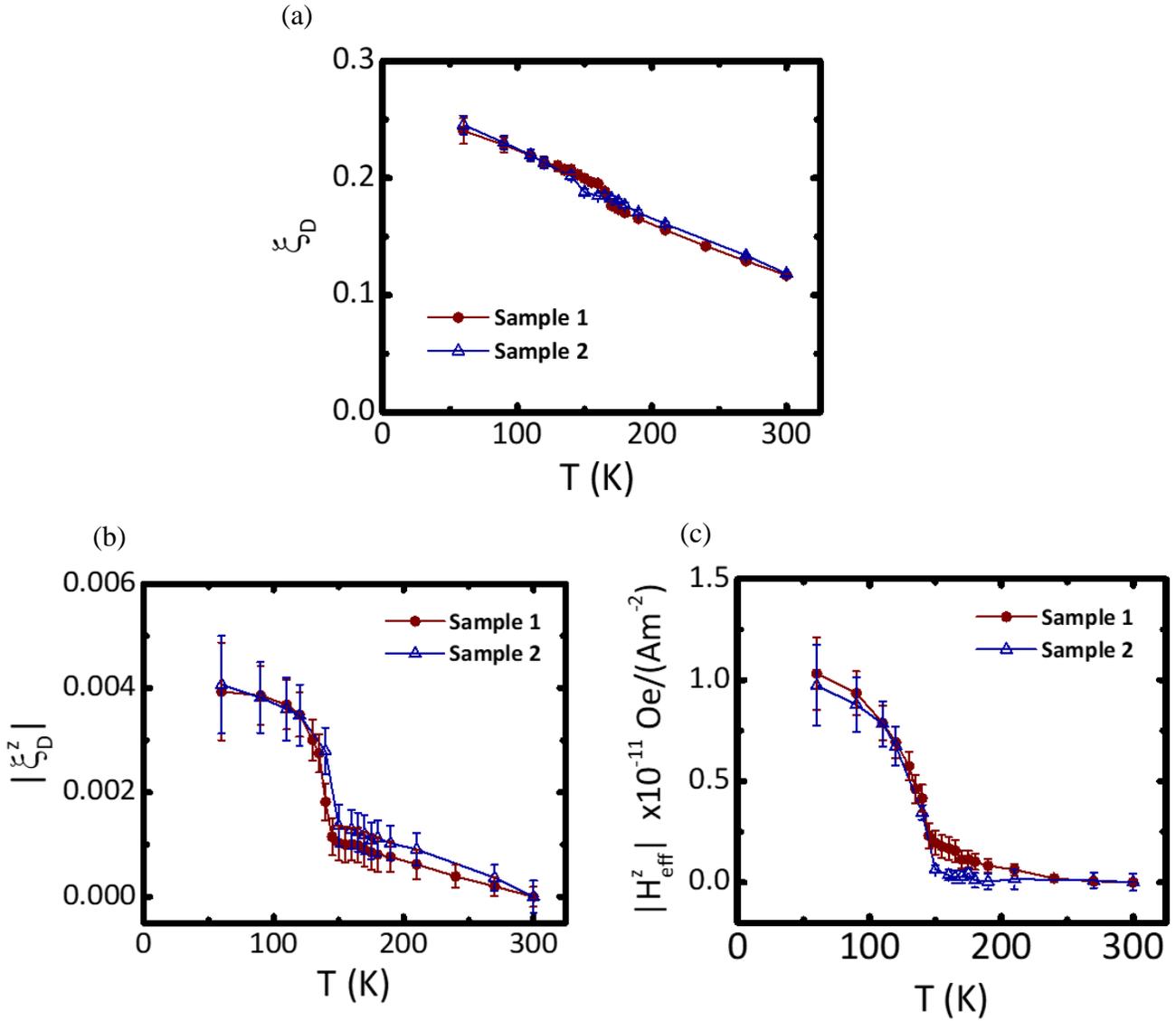

Figure S7: Comparison between two SRO(10)/Co(6) samples for the temperature dependence of (a) the SHE damping-like torque efficiency, (b) the z-type damping-like torque efficiency and (c) the z-type effective field.

In Fig. S7, we plot the efficiencies for different components of the spin torque vector, comparing the results from two microstrip samples (sample 1 is the one exhibited in the main text). As shown in Fig. S7(a), the T-dependent trend in the large SHE-induced damping-like spin torque efficiency is quantitatively very similar for both samples, apart from the details in the behavior in the vicinity of the SRO ferromagnetic transition. The much smaller amplitude of the temperature dependent features in the damping-like and field-like torques with symmetry indicative of a $\hat{s}_z$ spin current polarization and perpendicular effective field, plotted respectively in Fig. S7 (b) and (c) are also quite similar in amplitude, but again with small differences which we attribute to sample-to-sample differences in the magnetic domain configuration in the SRO microstrips.